\documentclass[12pt]{article}
\pdfoutput=1
\usepackage[nosort]{cite}

\usepackage{xcolor}

\usepackage{epsfig}
\usepackage{amsfonts}
\usepackage{amscd}
\usepackage{latexsym}
\usepackage{amsmath,amssymb}
\usepackage{verbatim}
\usepackage{setspace}
\usepackage{color}
\usepackage{fancyhdr}
\usepackage{cite}
\usepackage{hyperref}
\usepackage{tikz}
\usepackage{slashed}
\usepackage{multirow}
\usetikzlibrary{calc}
\usetikzlibrary{topaths}
\usetikzlibrary{decorations}
\usetikzlibrary{decorations.pathmorphing}
\usetikzlibrary{arrows,decorations.markings,cd}
\usetikzlibrary{calc,arrows,cd,decorations.markings,snakes}
\tikzset{
->-/.style args={#1rotate#2}{decoration={markings, mark=at position #1 with {\arrow[scale=1.5,rotate = #2 ]{stealth}}}, postaction={decorate}}
}
\usetikzlibrary{shapes.geometric}
\usetikzlibrary{knots}
\usepackage{tikz-3dplot}

\usepackage{draft}
\usepackage{hyperref}
\usepackage{graphicx,color,subfig}
\usepackage{cite}
\usepackage{mciteplus}
\usepackage{skak}
\usepackage{bbm}

\numberwithin{equation}{section}

\def\mm{\mathrm{m}}
\def\ww{\mathrm{w}}
\def\bZ{\mathbb{Z}}

\def\ot{{\mathcal{T}}}
\def\ls{\mathbf{a}}

\def\({\left(}
\def\){\right)}

\begin{document}

\begin{titlepage}

\title{Exactly Solvable 1+1d Chiral Lattice Gauge Theories}

\author{Sahand Seifnashri}

\address{School of Natural Sciences, Institute for Advanced Study,\\
1 Einstein Drive, Princeton, NJ 08540, USA}
\email{sahand@ias.edu}

\abstract{Using the modified Villain lattice Hamiltonian formulation of the 1+1d compact boson theory, we construct exactly solvable abelian chiral lattice gauge theories in two spacetime dimensions. As a concrete example, we derive an explicit quadratic lattice Hamiltonian for the ``34-50'' chiral gauge theory. We further show that $N$ copies of the modified Villain theory realize the $O(N,N;\bZ)$ T-duality transformations, which we then use to solve and analyze these lattice gauge theories.}

\end{titlepage}

\tableofcontents

\section{Introduction}

Constructing a lattice regularization of the Standard Model has been a longstanding challenge. The primary obstacle is the presence of chiral fermions, which are notoriously difficult to formulate on the lattice due to the Nielsen–Ninomiya theorem~\cite{Nielsen:1980rz}. The difficulty arises primarily from the anomaly of chiral fermions in even-dimensional spacetimes, for a review see~\cite{Kaplan:2009yg}.

However, the matter content of the Standard Model is special due to a delicate cancellation of anomalies. Thus, even if individual chiral fermions cannot be regularized on the lattice, it may still be possible to construct a lattice regularization of anomaly-free chiral gauge theories.

A widely used strategy for constructing a local lattice Hamiltonian for chiral gauge theories is to gap out the fermion doublers, or mirror fermions, while preserving the chiral, anomaly-free symmetries \cite{Eichten:1985ft, Poppitz:2010at, Chen:2012di}. In modern terms, this approach is formulated as \emph{symmetric mass generation}, which has been developed and studied in~\cite{Wen:2013ppa, Wang:2013yta, DeMarco:2017gcb, Wang:2018ugf, Tong:2021phe, Zeng:2022grc, Wang:2022ucy, Lu:2022qtc, Hasenfratz:2025lti}. However, maintaining control over these theories is difficult due to the strong dynamics of mirror fermions; for recent progress on this issue, see \cite{Mouland:2025ilu}. For alternative methods, see \cite{Kaplan:1992bt, Narayanan:1994gw, Luscher:1998du, Neuberger:2001nb, Bhattacharya:2006dc, Catterall:2023nww, Kaplan:2023pxd, Morikawa:2024zyd}.

In this work, we present a new approach that addresses this problem in two spacetime dimensions with abelian gauge groups. Specifically, we construct arbitrary 1+1d anomaly-free abelian chiral gauge theories on the lattice. Notably, our model is quadratic and solvable.

A prime example is the ``34-50'' U(1) chiral gauge theory. In this U(1) gauge theory, there are two Dirac fermions where the left-moving Weyl fermions have charges $3$ and $4$, and the right-moving Weyl fermions have charges $5$ and $0$. The pure gravitational anomaly vanishes because there are equal numbers of left-moving and right-moving fermions. The gauge anomaly also vanishes since
\ie
3^2 + 4^2 = 5^2 + 0^2\,.
\fe 

To construct anomaly-free chiral lattice gauge theories, we must find a mechanism for anomaly cancellation on the lattice. A crucial step is the lattice realization of \emph{global} symmetries with 't Hooft anomalies, prior to gauging them. Another essential step is gauging anomaly-free combinations of chiral symmetries. Both of these challenges have been successfully addressed in recent literature for abelian chiral symmetries in 1+1 dimensions.

The modified Villain formulation of compact boson CFT in 1+1d realizes the chiral symmetries of the theory exactly on the lattice \cite{Gross:1990ub,Sulejmanpasic:2019ytl,Gorantla:2021svj,Cheng:2022sgb,Fazza:2022fss}. Through bosonization, this construction yields a lattice realization of Dirac fermions and their exact chiral symmetries, thereby addressing the first challenge mentioned above.\footnote{See also \cite{Chatterjee:2024gje,Gioia:2025bhl}, where chiral anomalies are realized on the lattice through an enlarged, infinite-dimensional ``Onsager algebra'' symmetry. We do not pursue this approach here, as the anomaly cancellation mechanism for such symmetries is not understood.}

Recent progress in the study of 't Hooft anomalies of global symmetries on the lattice has clarified the computation of anomalies and consequently conditions for anomaly cancellation \cite{Else:2014vma,Kawagoe:2021gqi,Cheng:2022sgb,Seifnashri:2023dpa,Kapustin:2024rrm,Seifnashri:2025vhf,Bols:2025fmx,Pace:2025rfu,Kapustin:2025nju,Kawagoe:2025ldx,Tu:2025bqf,Shirley:2025yji,Seiberg:2025zqx,Feng:2025qgg,Czajka:2025mme,Seiberg:2026icc}. In particular, in \cite{Seifnashri:2023dpa} we developed a systematic framework for gauging anomaly-free symmetries directly on the lattice.

Combining these elements, we construct lattice chiral gauge theories in 1+1d, as proposed in \cite{Seifnashri:2023dpa}. A Euclidean formulation was subsequently worked out in \cite{Berkowitz:2023pnz}. Specifically, we construct quadratic lattice Hamiltonians for U(1) gauge theories coupled to $N$ Dirac fermions in 1+1 dimensions, where the U(1) symmetry is anomaly-free but chiral, acting differently on the left-moving and right-moving Weyl fermions. To achieve this, we begin by formulating a bosonized version of the theory, consisting of a U(1) gauge theory coupled to $N$ compact bosons, and then obtain the fermionic theory through fermionization.

An important new ingredient in this work is the realization of the T-duality group of $N$ compact bosons on the lattice. In particular, $N$ copies of the modified Villain Hamiltonian possess the full $O(N,N;\bZ)$ T-duality group, as we show explicitly for $N=2$ in Appendix \ref{app:T-duality}. In future work, we will discuss the T-duality transformations in more detail, and also discuss the lattice realization of the moduli spaces of the $T^N$ Sigma model on the lattice \cite{WIP}.

To ultimately formulate the Standard Model on the lattice, this construction needs to be generalized to both 3+1 dimensions and to non-abelian chiral symmetries. For discussions of lattice chiral symmetries in higher dimensions using Villain formulations, see \cite{Jacobson:2023cmr, Jacobson:2024hov, Peng:2025nfa, Xu:2024hyo, Fidkowski:2025rsq}. Realizing non-abelian chiral symmetries on the lattice remains a highly interesting open problem.\footnote{For a discussion of a non-abelian Villain lattice models, see \cite{Chen:2024ddr}.}

In Section~\ref{sec:mV}, we review the modified Villain model, discuss its properties, and solve the model. Section~\ref{sec:bos} presents our construction of abelian lattice chiral gauge theories and, in particular, the 34-50 theory. In Section~\ref{sec:schwinger}, we provide an exact solution to the case of U(1) gauge theory coupled to a single Dirac fermion and demonstrate its equivalence to the massless Schwinger model in the continuum limit. In Appendix~\ref{app:T-duality}, we derive and analyze the $O(2,2;\bZ)$ T-duality group of the two-flavor modified Villain lattice model. Finally, in Appendix \ref{app:fermionization}, we review the fermionization procedure used in Section~\ref{sec:bos} to construct the fermionic 34-50 theory.

\textit{Note added:} When this work was being finished, the work \cite{Thorngren:2026ydw} appeared, which also constructs lattice chiral gauge theories from the modified Villain model.

\section{The modified Villain Hamiltonian \label{sec:mV}}

Here, we review the modified Villain Hamiltonian \cite{Cheng:2022sgb,Fazza:2022fss} and its properties. The modified Villain model is a lattice regularization of the 1+1d $c=1$ compact boson CFT, and manifests its chiral symmetries as exact U(1) symmetries on the lattice.

The Hamiltonian on a periodic chain of $L$ sites is
\ie
H_\mathrm{mV} = \sum_{j=1}^L \left( \frac{1}{2R^2} p_j^2 + \frac{R^2}{2} \left( \tilde{p}_{j,j+1} + \frac{\phi_{j+1} - \phi_j}{2\pi} \right)^2 \right) \,, \label{mV}
\fe
with non-compact fields $\phi_j$ and $\tilde{\phi}_{j,j+1}$ respectively at site $j$ and link $(j,j+1)$. The conjugate momenta are $p_j$ and $\tilde{p}_{j,j+1}$ that satisfy the commutation relations
\begin{equation}
 [\phi_j , p_{j'}] =  [\tilde{\phi}_{j,j+1}, \tilde{p}_{j',j'+1}] =i \delta_{j,j'} \,. \label{ccr}
\end{equation}
In addition, we have Gauss's law constraints at sites and constraints at links that force $\tilde{\phi}_{j,j+1}$ to be $ 2\pi$-periodic and hence compact. Namely,
\begin{equation}
	e^{ 2\pi i p_j - i(\tilde{\phi}_{j,j+1}-\tilde{\phi}_{j-1,j})}=1 \qquad \text{and} \qquad e^{2\pi i\tilde{p}_{j,j+1}}=1 \,. \label{Gauss}
\end{equation}

\subsection{The chiral global symmetries}
The global symmetries that we are interested in here are the U(1)$_\mm$ momentum and U(1)$_\ww$ winding symmetries associated with the conserved charges
\ie
	Q_\mm = \sum_{j=1}^L q_j  \qquad \text{and} \qquad  Q_\ww = \sum_{j=1}^L  q_{j,j+1}\,,
\fe
where
\ie
	q_j = p_j - \frac{\tilde{\phi}_{j,j+1} -\tilde{\phi}_{j-1,j}}{2\pi} \qquad \text{and} \qquad q_{j,j+1} = \tilde{p}_{j,j+1}\,.
\fe
Because of Gauss's constraints \eqref{Gauss}, the local charges $q_j$ and $q_{j,j+1}$ are integer, and therefore the momentum and winding charges $Q_\mm$ and $Q_\ww$ are quantized. Moreover, even though $q_j$ and $q_{j,j+1}$ are not gauge invariant, the global charges are gauge invariant since they can be rewritten as
\ie
	Q_\mm = \sum_{j=1}^L p_j  \qquad \text{and} \qquad  Q_\ww = \sum_{j=1}^L  \left( \tilde{p}_{j,j+1} + \frac{\phi_{j+1} - \phi_j}{2\pi} \right)\,,
\fe
which manifests their gauge-invariance but obscures their quantization.

The momentum and winding U(1) symmetries are separately free of any 't Hooft anomaly. However, as explained in \cite{Cheng:2022sgb}, there is a mixed 't Hooft anomaly between them associated with the commutation relation between the local charges
\ie
	{} [q_j , q_{j,j+1}] = [q_{j-1,j} , q_j] = \frac{-i}{2\pi} \,,
\fe 
which resembles the Schwinger term in the continuum theory. For a different and quantitative derivation of the anomaly, see \cite{Seifnashri:2023dpa}.

In other words, the U(1) symmetries associated with the conserved charges $Q_\mathrm{L}=Q_\mm + Q_\ww$ and  $Q_\mathrm{R}=Q_\mm - Q_\ww$ are anomalous and generate the chiral U(1)$_\mathrm{L}$ and U(1)$_\mathrm{R}$ symmetries where\footnote{At a generic radius $R$, the charges $Q_\mathrm{L}$ and $Q_\mathrm{R}$ are not purely left- and right-moving. We nevertheless label them as L and R, since their 't Hooft anomalies are chiral, namely $k_\mathrm{L} = 1$ and $k_\mathrm{R} = -1$.}
\ie
	\mathrm{U}(1)_\mm \times \mathrm{U}(1)_\ww = \frac{\mathrm{U}(1)_\mathrm{L} \times \mathrm{U}(1)_\mathrm{R}}{\mathbb{Z}_2} \,.
\fe
The $\bZ_2$ quotient corresponds to the identification $e^{\pi i Q_\mathrm{L}} = e^{\pi i Q_\mathrm{R}}$.

\subsection{Solving the model \label{sec:mv.sol}}

The modified Villain Hamiltonian is quadratic and can be solved exactly. Here, we solve it explicitly and find the Hamiltonian's spectrum, matching it with the continuum result.

We begin by defining the variables $\Phi_1, \Phi_2, \cdots, \Phi_{2L}$ as
\ie
	\Phi_{2j} = \frac{p_j}{R}\,, \qquad \Phi_{2j+1} = R \left( \tilde{p}_{j,j+1} + \frac{\phi_{j+1} - \phi_{j}}{2\pi} \right) \qquad \text{for } j =1, \cdots, L \,,
\fe
which form a periodic chain of $2L$ sites. They satisfy the commutation relations
\ie
	{[\Phi_\ell , \Phi_{\ell+1} ]} =\frac{i}{2\pi} \,,
\fe
while $[\Phi_\ell , \Phi_{\ell'}] = 0$ for $\ell - \ell' \neq \pm 1$. We define
\ie
	\Pi_k = \frac{1}{\sqrt{2L}} \sum_{\ell=1}^{2L} e^{\frac{2\pi i}{2L} \ell k} \Phi_\ell \,,
\fe
where $\Pi_{-k} = \Pi_k^\dagger$ and they satisfy
\ie
	{ \left[ \Pi_k , \Pi_{k'}^\dagger \right] } = \frac{\sin(\frac{2\pi k}{2L})}{\pi} \delta_{k,k'} \,.
\fe

There are two zero modes
\ie
	\Pi_0 = \frac{1}{\sqrt{2L}} \left( \frac{Q_\mm}{R} + R Q_\ww \right) \qquad \text{and} \qquad \Pi_L = \frac{1}{\sqrt{2L}} \left( \frac{Q_\mm}{R} - R Q_\ww \right) \,,
\fe
and non-zero (oscillatory) modes with energy $\omega_k = \sin(\frac{2\pi k}{2L})/\pi$ are associated with the annihilation operators $a_k = \sqrt{\omega_k} \, \Pi_k$ for $k =1,2, \cdots, L-1$ satisfying $[a_k ,a_{k'}^\dagger] = \delta_{k,k'}$. The Hamiltonian can be rewritten as\footnote{The $\Phi_\ell$ variables relate the modified Villain model with the ``staggered boson'' model of \cite{Berenstein:2023tru, Berenstein:2023ric}. The difference between them is in the zero modes $\Pi_0$ and $\Pi_L$. In the staggered boson model, the zero modes are not a priori quantized, and their conjugate variables are missing. In the modified Villain model, the zero modes are quantized, and their conjugate variables correspond to gauge-invariant operators $e^{i \phi_j}$ and $e^{i \tilde{\phi}_{j,j+1}}$.}
\ie
	H_\mathrm{mV} &= \frac{1}{2} \sum_{\ell=0}^{2L-1}  \Phi_\ell^2 = \frac{1}{2} \sum_{k=0}^{2L-1} \Pi_k^\dagger \Pi_k  \\
	&=  \frac{1}{4L}\left(\frac{Q_\mm}{R} + R Q_\ww \right)^2 + \frac{1}{4L}\left(\frac{Q_\mm}{R} - R Q_\ww \right)^2 + \sum_{k=1}^{L-1} \omega_k \left(a_k^\dagger a_k + \frac{1}{2} \right).
\fe

In the continuum limit $L \to \infty$, we focus on the low-energy spectrum and neglect high-energy modes. In this regime, the Hamiltonian decomposes into left/right moving modes, $H_\mathrm{mV} \sim \frac{1}{L} \left( L_0 + \bar{L}_0 \right)$, with
\ie
	L_0 = \frac{1}{4}\left(\frac{Q_\mm}{R} + R Q_\ww \right)^2 + \sum_{n=1}^\infty n \left(a_n^\dagger a_n + \frac{1}{2} \right) \,,\\
	\bar{L}_0 = \frac{1}{4}\left(\frac{Q_\mm}{R} - R Q_\ww \right)^2 + \sum_{\bar{n}=1}^\infty \bar{n} \left(a_{\bar{n}}^\dagger a_{\bar{n}} + \frac{1}{2} \right) \,.
\fe
Here we have used the low-momentum approximation $\omega_k \sim k/L$ and $\omega_k \sim (L-k)/L$, valid for $n=k \ll L$ or $\bar{n} = L-k \ll L$, and have discarded high-energy modes. The spectrum of $ L_0 + \bar{L}_0$ precisely reproduces that of the 1+1d compact boson CFT at radius R. This analysis further shows that the correctly normalized, dimensionful Hamiltonian is
\ie
	H(\ls) = \frac{2\pi}{\ls} H_\mathrm{mV} \,, \label{dim.ful.H}
\fe
where $\ls$ denotes the lattice spacing.

\section{Chiral lattice gauge theory Hamiltonians \label{sec:bos}}

We begin with the $N$-flavor modified Villain Hamiltonian
\ie
H_\mathrm{mV} = \sum_{I=1}^N \sum_{j=1}^L \left( \frac{1}{2R^2} \left(p_j^{(I)}\right)^2 + \frac{R^2}{2} \left( \tilde{p}_{j,j+1}^{(I)} + \frac{\phi_{j+1}^{(I)} - \phi_j^{(I)}}{2\pi} \right)^2 \right) \,, \label{mV}
\fe
where $p_{j}^{(I)}$ and $\tilde{p}_{j,j+1}^{(I)}$ are the conjugate momenta of $\phi_j^{(I)}$ and $\tilde{\phi}_{j,j+1}^{(I)}$, respectively, as in \eqref{ccr}. The Gauss law constraints are
\ie
	\exp{\left(2\pi i \, q_j^{(I)}\right)} = 1  \qquad \text{and} \qquad \exp{\left( 2\pi i \, q_{j,j+1}^{(I)} \right)} = 1 \,,\label{Gauss.law}
\fe
where
\ie
	q_j^{(I)} = p_j^{(I)} - \frac{\tilde{\phi}_{j,j+1}^{(I)} -\tilde{\phi}^{(I)}_{j-1,j}}{2\pi} \qquad \text{and} \qquad q_{j,j+1}^{(I)} = \tilde{p}_{j,j+1}^{(I)} \,.
\fe

We consider the global U(1) symmetry associated with the quantized conserved charge
\ie
	Q = \sum_{I=1}^N \Big( n_\mm^{(I)} Q_\mm^{(I)} + n_\ww^{(I)} Q_\ww^{(I)} \Big)\,, \label{conserved.charge}
\fe
with $n_\mm^{(I)}, n_\ww^{(I)} \in \bZ$ and
\ie
	Q_\mm^{(I)} = \sum_{j=1}^L q_j^{(I)} \qquad \text{and} \qquad Q_\ww^{(I)} = \sum_{j=1}^L  q_{j,j+1}^{(I)} \,.
\fe

Using the commutation relation between the local charges, namely
\ie
	{} [q_j^{(I)} , q_{j,j+1}^{(I)}] = [q_{j-1,j}^{(I)} , q_j^{(I)}] = \frac{-i}{2\pi} \,,
\fe
one can compute the 't Hooft anomaly of the U(1) symmetry generated by $Q$ \cite{Cheng:2022sgb}. The U(1) symmetry is anomaly-free precisely when
\ie
	\left[\sum_I n_\mm^{(I)} q_{j-1}^{(I)} + n_\ww^{(I)} q_{j-1,j}^{(I)} \, , \sum_I n_\mm^{(I)} q_j^{(I)} + n_\ww^{(I)} q_{j,j+1}^{(I)} \right] = \frac{-i}{2\pi} \sum_{I=1}^N n_\mm^{(I)} n_\ww^{(I)} = 0 \,. \label{anomaly-free}
\fe

Following \cite{Seifnashri:2023dpa}, we now discuss how to gauge such chiral non-on-site yet anomaly-free U(1) symmetries on the lattice. We derive two equivalent descriptions of such lattice chiral gauge theories. In the first description, the gauge fields couple through the Gauss law constraints. While in the second presentation, they couple directly through the Hamiltonian.

\subsection{First presentation}

The gauging procedure consists of two steps: coupling to gauge fields and imposing Gauss's law constraints. The U(1) gauge fields consist of compact fields $a_j$ at sites and their conjugate variable $E_j$, the electric field, satisfying
\ie
	{[a_j , E_{j'}]} = i \delta_{j,j'} \,.
\fe	
We take $a_j$ to be $2\pi$-periodic implying that the electric field $E_j$ is integer valued:
\ie
	\exp{\left( 2\pi i E_j \right)} =1 \,. \label{electric.field}
\fe

\subsubsection*{I. Coupling to background gauge fields}
In the first presentation, coupling to gauge fields is implemented by modifying the constraints in \eqref{Gauss.law} as described in \cite{Cheng:2022sgb, Seifnashri:2023dpa}. Inserting a defect for the symmetry transformation $e^{-i\theta Q}$, with the conserved charge $Q$ defined in \eqref{conserved.charge}, at site $j_*$ is equivalent to conjugating the system by the unitary operator
\ie
	\exp{ \left[ - i \theta \sum_{j<j_*} \sum_{I=1}^N \left( n_\mm^{(I)} q_j^{(I)} + n_\ww^{(I)} q_{j,j+1}^{(I)}  \right) \right]} \,.
\fe
This similarity transformation leaves the Hamiltonian invariant, since it commutes with it, and instead modifies only Gauss’s law constraints. Namely, it modifies Gauss's laws at site $j_*$ and on link $(j_*-1,j_*)$ to\footnote{The first constraint is a modified Gauss's law constraint associated with the $\bZ$ gauge fields of the modified Villain model, and the second one corresponds to changing the periodicity of the electric fields $\tilde{\phi}_{j,j+1}$.}
\ie
	e^{2\pi i q_{j_*}^{(I)}} = e^{i n_\ww^{(I)} \theta} \qquad \text{and} \qquad e^{2\pi i q_{j_*-1,j_*}^{(I)}} = e^{i n_\mm^{(I)} \theta} \,.
\fe

\subsubsection*{II. Making gauge fields dynamical}
The gauge field configuration $a_j$ corresponds to inserting an $a_j \in \mathrm{U}(1)$ defect at site $j$ for all $j$. As a result, Gauss's law at site $j$ and flavor $I$ is modified to
\ie
	e^{2\pi i q_j^{(I)}} = e^{i n_\ww^{(I)} a_j} \qquad \text{and} \qquad e^{2\pi i q_{j-1,j}^{(I)}} = e^{i n_\mm^{(I)} a_j} \,. \label{modified.g.l}
\fe
Making the gauge field dynamical amounts to imposing Gauss's law constraints for the U(1) gauge fields on each link $(j,j+1)$, given by
\ie
	G_{j,j+1} = \sum_{I=1}^N \big( n_\mm^{(I)} q_j^{(I)} + n_\ww^{(I)} q_{j,j+1}^{(I)} \big) + E_{j+1} - E_j = 0 \,. \label{u1.g.l}
\fe
The anomaly-free condition, $\sum_I n_\mm^{(I)} n_\ww^{(I)} = 0$, guarantees that Gauss's law constraint commutes at different sites, namely $[G_{j,j+1}, G_{j',j'+1}]=0$.

The Hamiltonian of the gauged theory includes a kinetic term for the electric field:
\ie
	H_\mathrm{gauged} =&\; \frac{e^2}{2} \sum_{j=1}^L  \left( E_j + \frac{\theta}{2\pi} - \sum_{I=1}^N \frac{n_\ww^{(I)} \phi_j^{(I)} + n_\mm^{(I)} \tilde{\phi}_{j-1,j}^{(I)}  }{2\pi} \right)^2 \\ &+  \sum_{I=1}^N \sum_{j=1}^L \left( \frac{1}{2R^2} \left(p_j^{(I)}\right)^2 + \frac{R^2}{2} \left( \tilde{p}_{j,j+1}^{(I)} + \frac{\phi_{j+1}^{(I)} - \phi_j^{(I)}}{2\pi} \right)^2 \right) \,. \label{H.gauged}
\fe	
Here, $\theta \sim \theta + 2\pi$ is the theta angle and $e$ is the dimensionless electric coupling. Its relation to the dimensionful coupling $e_\mathrm{cont}$ of the continuum theory is
\ie
	e = \frac{e_\mathrm{cont} \, \ls}{\sqrt{2\pi}}\,. \label{dim.ful.coupling}
\fe 
Here, $\ls$ is the lattice spacing introduced in \eqref{dim.ful.H}, such that the dimensionful Hamiltonian is $H_\mathrm{gauged}(\ls) = ({2\pi}/{\ls}) H_\mathrm{gauged}$. In summary, the lattice gauge theory is defined by the Hamiltonian \eqref{H.gauged} and is subject to Gauss law constraints in \eqref{electric.field}, \eqref{modified.g.l}, and \eqref{u1.g.l}.

\subsection{Second presentation}

In the second presentation, which takes a more familiar form, we keep the constraint \eqref{Gauss.law} unchanged and couple the gauge fields directly through the Hamiltonian. To find this presentation, we start with the first presentation and apply the similarity transformation given by the unitary operator
\ie
	\exp{\left( -i\sum_{I,j} \left(\frac{n_\ww^{(I)}}{2\pi} \phi^{(I)}_j + \frac{n_\mm^{(I)}}{2\pi} \tilde{\phi}^{(I)}_{j-1,j} \right) a_j \right)} \,.
\fe
Doing such, the gauged Hamiltonian becomes
\ie
	H_\mathrm{gauged} =&\; \frac{e^2}{2} \sum_{j=1}^L \left( E_j + \frac{\theta}{2\pi} \right)^2 +  \sum_{I=1}^N \sum_{j=1}^L  \frac{1}{2R^2} \left(p_j^{(I)} + \frac{n_\ww^{(I)}}{2\pi} a_j \right)^2 \\ & + \sum_{I=1}^N \sum_{j=1}^L  \frac{R^2}{2} \left( \tilde{p}_{j,j+1}^{(I)} + \frac{n_\mm^{(I)}}{2\pi} a_{j+1} + \frac{\phi_{j+1}^{(I)} - \phi_j^{(I)}}{2\pi} \right)^2  \,,
\fe
Gauss's law for the U(1) gauging becomes
\ie
	G_{j,j+1} = \sum_{I=1}^N \left( n_\mm^{(I)} p_j^{(I)} + n_\ww^{(I)}\left( \tilde{p}_{j,j+1}^{(I)} + \frac{\phi_{j+1}^{(I)} - \phi_j^{(I)}}{2\pi} \right) \right) + E_{j+1} - E_j = 0 \,, \label{u1.gauss}
\fe
while the original constraints \eqref{Gauss.law} remain unchanged:
\begin{equation}
	\exp{\left(  2\pi i p_j^{(I)} - i\big(\tilde{\phi}_{j,j+1}^{(I)}-\tilde{\phi}_{j-1,j}^{(I)}\big) \right)}=1 \, ,\qquad \exp{ \left( 2\pi i\tilde{p}_{j,j+1}^{(I)} \right)}=1 \,, \label{2.gauss.law}
\end{equation}
and the quantization of the electric field changes from \eqref{electric.field} to
\ie
	e^{2\pi i E_j} = \exp{\left( -i\sum_{I=1}^N \frac{n_\ww^{(I)}}{2\pi} \phi^{(I)}_j + \frac{n_\mm^{(I)}}{2\pi} \tilde{\phi}^{(I)}_{j-1,j} \right)} \,. \label{3.gauss.law}
\fe

Again, it is easy to check that the Hamiltonian is gauge invariant and commutes with the constraints \eqref{u1.gauss}, \eqref{2.gauss.law}, and \eqref{3.gauss.law}. Moreover, all the constraints commute mutually with each other and at different sites, given the anomaly-free condition \eqref{anomaly-free}.

\subsection{Warm-up: $N=1$ and the massless Schwinger model \label{sec:schwinger}}

As a warm-up and a consistency check of our construction, we begin with the $N=1$ case and solve it exactly. For $R=1/\sqrt{2}$, $n_\mm=0$, and $n_\ww=1$, the resulting (non-chiral) U(1) gauge theory is expected to be equivalent to the massless Schwinger model by bosonization \cite{Coleman:1975pw,Coleman:1976uz}. Indeed, we will verify this equivalence explicitly at the end of this subsection.\footnote{For a recent lattice study of this model using staggered fermions, see \cite{Dempsey:2022nys}.}

We start from a single compact boson and gauge a $\mathbb{Z}_{n_\ww}$ cover of its winding symmetry. This corresponds to setting $N=1$ and $n_\mm=0$. The resulting Hamiltonian takes the form
\ie
	H_\mathrm{Schwinger} = \sum_{j=1}^L \left( \frac{e^2}{2} \left( E_j + \frac{\theta}{2\pi} -  \frac{n_\ww \phi_j   }{2\pi} \right)^2 +  \frac{1}{2R^2} p_j^2 + \frac{R^2}{2} \left( \tilde{p}_{j,j+1} +  \frac{\phi_{j+1} - \phi_j }{2\pi} \right)^2 \right) \,, 
\fe
subject to the constraints
\ie
	\exp{ \left( 2\pi i\left( p_j- \frac{\tilde{\phi}_{j,j+1} -\tilde{\phi}_{j-1,j}}{2\pi}  \right) \right)} &= \exp{\left( i n_\ww a_j \right)} \,, & \quad \exp{ \left( 2\pi i \tilde{p}_{j,j+1} \right) } &= 1 \,, \\
	 n_\ww \tilde{p}_{j,j+1} + E_{j+1} - E_j &= 0 \,,&
	\exp{\left( 2\pi i E_j \right)} &= 1 \,.
\fe

Doing a similarity transformation given by
\ie
	\exp{ \left( \frac{2\pi i}{n_\ww}\sum_{j=1}^L E_j q_j \right) } \,,
\fe
the Hamiltonian of the bosonized charge-$n_\ww$ Schwinger model becomes
\ie
	H_{e,\mathrm{n_\ww}} = \frac{e^2}{2} \sum_{j=1}^L  \left( \frac{\theta}{2\pi} -  \frac{n_\ww \phi_j  }{2\pi} \right)^2 +   \sum_{j=1}^L \frac{1}{2R^2} p_j^2 + \sum_{j=1}^L  \frac{R^2}{2} \left( \tilde{p}_{j,j+1} + \frac{\phi_{j+1}- \phi_j}{2\pi} \right)^2  \,,
\fe
with constraints
\ie
	 \exp{\left( i n_\ww a_j \right)} &= 1 \,, & \qquad \exp{ \left( 2\pi i \tilde{p}_{j,j+1} \right) } &= \exp{\left( \frac{2\pi i}{n_\ww} \left( E_{j+1}-E_j \right) \right)} \,, \\
	 n_\ww \tilde{p}_{j,j+1} &= 0 \,,&
	\exp{\left( 2\pi i E_j \right)} &= 1 \,.
\fe
The constraints completely trivialize $\tilde{\phi},\tilde{p}$ degrees of freedom on the links and decouple $a_j,E_j$ as a TQFT that is the $\mathbb{Z}_{n_\ww}$ gauge theory associated with the topological local operator $\exp{\left( \frac{2\pi i}{n_\ww} E_j \right)} = \exp{\left( \frac{2\pi i}{n_\ww} E_{j+1} \right)}$. Thus, we are left with a $\mathbb{Z}_{n_\ww}$ gauge theory coupled to the charge $1$ Schwinger model described by
\ie
	H_{\mathrm{n_\ww}e} = \frac{n_\ww^2 e^2 }{8\pi^2} \sum_{j=1}^L \phi_j^2 +   \frac{1}{2R^2} \sum_{j=1}^L  p_j^2 + \frac{R^2}{2}  \sum_{j=1}^L  \left( \frac{\phi_{j+1}- \phi_j}{2\pi} \right)^2  \,. 
\fe

Thus, we find that the 1+1d charge $n$ Schwinger model with electric coupling $e$ is the same theory as the charge $n'=1$ Schwinger model at electric coupling $e'=n e$ tensored with $\mathbb{Z}_{n}$ gauge theory.

To solve the model, define
\ie
	\phi_j &= \sum_{k=0}^{L-1}  \frac{1}{\sqrt{2 L \omega_k} } \left( e^{-2\pi i \frac{jk}{L}} a_k^\dagger + e^{2\pi i \frac{jk}{L}} a_k\right) ,  & p_j &= \sum_{k=0}^{L-1} \sqrt{\frac{\omega_k}{2L}} \left( i e^{-2\pi i \frac{jk}{L}} a_k^\dagger - i e^{2\pi i \frac{jk}{L}} a_k \right) \,,
\fe
where $[a_k,a_{k'}^\dagger] = \delta_{k,k'}$, and substitute it in the Hamiltonian to find
\ie
	H_{\mathrm{n_\ww}e} = \; & \frac{n_\ww^2 e^2 +2R^2}{8\pi^2}  \sum_{k=0}^{L-1}  \frac{(a_k^\dagger+a_{-k})( a_{-k}^\dagger + a_k )}{2\omega_k} \\ & +  \frac{1}{2R^2} \sum_{k=0}^{L-1} \omega_k \frac{(ia_k^\dagger-ia_{-k})( ia_{-k}^\dagger - ia_k )}{2} \\ & - \frac{R^2}{4\pi^2}  \sum_{k=0}^{L-1} \cos\left(\frac{2\pi  k}{L} \right) \frac{(a_k^\dagger+a_{-k})(a_{-k}^\dagger + a_k)}{2\omega_k} \\
	=\; & \sum_{k=0}^{L-1} \frac{\omega_k}{R^2} \left(a_k^\dagger a_k + \frac{1}{2} \right) \,, \qquad \omega_k = \frac{R}{2\pi}\sqrt{n_\ww^2 e^2 +2R^2 \left(1-\cos\left(\frac{2\pi k}{L} \right) \right)} \,.
\fe
Incorporating the lattice spacing $\ls$ defined in \eqref{dim.ful.coupling}, we obtain the dispersion relation
\ie
	E_k = \frac{2\pi}{\ls} \omega_k = \sqrt{\frac{n_\ww^2 e^2 }{ R^2 \ls^2}+\frac{2}{\ls^2} \left(1-\cos\left(\frac{2\pi k}{L} \right) \right)} = \sqrt{\frac{n_\ww^2 e^2 }{R^2 \ls^2}+\left(\frac{2\pi k}{L \ls} \right)^2 + \cdots } ~.
\fe

In the continuum limit, this dispersion relation describes a free massive boson with mass
\ie
	M = \frac{n_\ww \, e_\mathrm{cont}}{\sqrt{2\pi} R}\,,
\fe
where $e_\mathrm{cont}$ is the continuum, dimensionful electric coupling constant. Indeed, setting $R=1/\sqrt{2}$ and $n_\ww=1$, one recovers the Schwinger boson mass, $M = e_\mathrm{cont} / \sqrt{\pi}$.

\subsection{The bosonized 34-50 theory}

Here we discuss and analyze an $N=2$ chiral lattice gauge theory, which is related to the 34-50 model by bosonization. To find the bosonized model, let us first review some details of bosonization for the $N=1$ case.

\subsubsection*{Review of bosonization}
The Dirac fermion theory is related by bosonization to the compact boson theory at radius $R=\sqrt{2}$, or equivalently at the T-dual radius $R=\frac{1}{\sqrt{2}}$. For a review, see \cite{Tachikawa2018Topological, Karch:2019lnn}. The chiral global symmetry of the Dirac theory is
\ie
	G_\mathrm{Dirac} = \mathrm{U}(1)_\mathrm{L} \times \mathrm{U}(1)_\mathrm{R} = \frac{\mathrm{U}(1)_\mathrm{V} \times \mathrm{U}(1)_\mathrm{A}}{\bZ_2^\mathrm{diag}} \,,
\fe
where $Q_\mathrm{V} = Q_\mathrm{L} + Q_\mathrm{R}$ and $Q_\mathrm{A} = Q_\mathrm{L} - Q_\mathrm{R}$. The symmetries of the bosonic theory at radius $R=\sqrt{2}$ is
\ie
	G_{\sqrt{2}} = \mathrm{U}(1)_\mathrm{m} \times \mathrm{U}(1)_\mathrm{w} = \mathrm{U}(1)_\mathrm{V} \times \frac{\mathrm{U}(1)_\mathrm{A}}{\bZ_2^\mathrm{A}} \,,
\fe
where $Q_\mathrm{V} = Q_\mathrm{m}$ and $Q_\mathrm{A} = 2Q_\mathrm{w}$. At the T-dual radius, the global symmetry is 
\ie
	G_{\frac{1}{\sqrt{2}}} = \mathrm{U}(1)_\mathrm{m} \times \mathrm{U}(1)_\mathrm{w} = \frac{\mathrm{U}(1)_\mathrm{V}}{\bZ_2^\mathrm{V}} \times \mathrm{U}(1)_\mathrm{A} \,,
\fe
where $Q_\mathrm{V} = 2Q_\mathrm{m}$ and $Q_\mathrm{A} = Q_\mathrm{w}$. In summary, the symmetry charges of the fermionic and bosonic theories are related as
\ie
	(Q_\mathrm{L}\,,~ Q_\mathrm{R}) = \begin{cases} \left(\frac{1}{2}Q_\mm + Q_\ww\,, ~\frac{1}{2}Q_\mm - Q_\ww \right) ~~&\text{for } R=\sqrt{2} \\ 
	\left(Q_\mm + \frac{1}{2}Q_\ww\,, ~Q_\mm - \frac{1}{2}Q_\ww \right) &\text{for } R=\frac{1}{\sqrt{2}}
	\end{cases} \,.
\fe
Note that the fermionic theory is obtained by \emph{fermionic} gauging of the $\bZ_2^\mm$ (or $\bZ_2^\ww$) symmetry of the bosonic theory at radius $R=\sqrt{2}$ (or $R=\frac{1}{\sqrt{2}}$).

One can repeat this story for multiple copies and obtain a duality between $N$ Dirac fermions and $N$ compact bosons at radius $R=\frac{1}{\sqrt{2}}$ (or $R=\sqrt{2}$). In that case, the bosonization map involves gauging a $\bZ_2^N$ symmetry on each side. In the bosonic theory, one must do fermionic gauging of the winding symmetry $\prod_{I=1}^N \bZ_2^{\ww_I}$, and on the fermionic side, one gauges $\prod_{I=1}^N \bZ_2^{\mathrm{V}_I}$.

Finally, consider two Dirac fermions and the anomaly-free, yet chiral, U(1) symmetry with the charge assignment 34-50. The 34-50 global symmetry is generated by the conserved quantized charge
\ie
	Q = 3 Q_\mathrm{L}^{(1)} + 4 Q_\mathrm{L}^{(2)} + 5 Q_\mathrm{R}^{(1)} = \begin{cases}
	4 Q_\mathrm{m}^{(1)} + 2 Q_\mathrm{m}^{(2)} - 2 Q_\mathrm{w}^{(1)} - 4 Q_\mathrm{w}^{(2)} ~&\text{for } R=\sqrt{2} \\ 
	8 Q_\mathrm{m}^{(1)} + 4 Q_\mathrm{m}^{(2)} - Q_\mathrm{w}^{(1)} + 2 Q_\mathrm{w}^{(2)} 	 &\text{for } R=\frac{1}{\sqrt{2}}
	\end{cases}  \,. \label{34-50-charge}
\fe
Therefore, the fermionic 34-50 chiral gauge theory is related by bosonization to two copies of compact boson theory at radius $R=\frac{1}{\sqrt{2}}$ and with the charge assignment $(n_\mm^{(1)}, n_\mm^{(2)}, n_\ww^{(1)}, n_\ww^{(2)}) = (8,4,-1,2)$. The fermionization involves fermionic gauging of $\bZ_2^{\ww_1} \times \bZ_2^{\ww_2}$. However, the symmetry $\bZ_2^{\ww_1}$ is already gauged since it is a subgroup of the gauged symmetry generated by \eqref{34-50-charge}, i.e., $e^{i \pi Q} = e^{i \pi Q_\ww^{(1)}}$. Thus, in this case, the fermionization involves only the fermionic gauging of $\bZ_2^{\ww_2}$.

\subsubsection*{The $(8,4,-1,2)$ model}

Setting $N=2$, $R=\frac{1}{\sqrt{2}}$, and $(n_\mm^{(1)}, n_\mm^{(2)}, n_\ww^{(1)}, n_\ww^{(2)}) = (8,4,-1,2)$ we find the bosonization of the 34-50 model, as explained above. The Hamiltonian for this lattice chiral gauge theory is
\ie
	H_{8,4,-1,2} =&\; \frac{e^2}{2} \sum_{j=1}^L  \left( E_j + \frac{\theta}{2\pi} - \frac{  8 \tilde{\phi}_{j-1,j}^{(1)} +4\tilde{\phi}_{j-1,j}^{(2)} - \phi_j^{(1)} +2 \phi_j^{(2)} }{2\pi} \right)^2 \\ &+   \sum_{j=1}^L \sum_{I=1}^2 \left(  \left(p_j^{(I)}\right)^2 + \frac{1}{4} \left( \tilde{p}_{j,j+1}^{(I)} + \frac{\phi_{j+1}^{(I)} - \phi_j^{(I)}}{2\pi} \right)^2 \right) \,, 
\fe
with constraints
\ie
	\exp{ \left( 2\pi i q_j^{(1)}  \right)} &= \exp{\left( - i  a_j \right)} \,, & \quad \exp{ \left( 2\pi i q_{j-1,j}^{(1)} \right) } &= \exp{\left(  8 i a_j \right)} \,, \\
	\exp{ \left( 2\pi i q_j^{(2)}  \right)} &= \exp{\left( 2 i  a_j \right)} \,, & \quad \exp{ \left( 2\pi i q_{j-1,j}^{(2)} \right) } &= \exp{\left(  4 i a_j \right)} \,, \\
	\exp{\left( 2\pi i E_j \right)} &= 1 \,,&
	8 q_j^{(1)} + 4 q_j^{(2)}  - q^{(1)}_{j,j+1} +2 q^{(2)}_{j,j+1} &= E_j -E_{j+1}  \,.
\fe

To solve the model, we perform a T-duality transformation $\ot_\mathcal{M}$ associated with the $O(2,2;\bZ)$ matrix
\ie
\mathcal{M} = \begin{pmatrix}
-1&2&8&4\\
0&-1&-4&0\\
0&0&-1&0\\
0&0&-2&-1
\end{pmatrix} \,.
\fe
The action of $\ot_\mathcal{M}$ on local operators, discussed in Appendix \ref{app:T-duality}, is
\ie
\ot_\mathcal{M}: ~ \begin{split} {\phi}_{j}^{(1)} &\mapsto -{\phi}_{j}^{(1)} -2 {\phi}_{j}^{(2)} -4 \tilde{\phi}^{(2)}_{j-1,j} +8\tilde{\phi}^{(1)}_{j,j+1}  \,, \;\qquad\qquad\qquad p_j^{(1)} \mapsto -p_j^{(1)} \,,  \\
 {\phi}_{j}^{(2)} &\mapsto -{\phi}_{j}^{(2)} + 4\tilde{\phi}^{(1)}_{j,j+1}  \,, ~\;\qquad\qquad\qquad\qquad\qquad p_j^{(2)} \mapsto -p_j^{(2)} +2p_j^{(1)} \,,  \\
\tilde{\phi}_{j,j+1}^{(1)} &\mapsto -\tilde{\phi}_{j,j+1}^{(1)}  \,, ~\qquad\qquad\, \tilde{p}_{j,j+1}^{(1)} \mapsto -\tilde{p}_{j,j+1}^{(1)} -2 \tilde{p}_{j,j+1}^{(2)} -4q^{(2)}_{j} + 8 q^{(1)}_{j+1} \,, \\
\tilde{\phi}_{j,j+1}^{(2)} &\mapsto -\tilde{\phi}_{j,j+1}^{(2)} + 2\tilde{\phi}_{j,j+1}^{(1)}  \,, \qquad\qquad\qquad\qquad \tilde{p}_{j,j+1}^{(2)} \mapsto -\tilde{p}_{j,j+1}^{(2)} +4 q^{(1)}_{j+1} \,. 
\end{split} 
\fe

Applying the T-duality transformation $\ot_\mathcal{M}$, the chiral gauge theory Hamiltonian becomes
\ie
	H_{8,4,-1,2} =&\; \frac{e^2}{2} \sum_{j=1}^L  \left( E_j + \frac{\theta}{2\pi} - \frac{  \phi_j^{(1)} }{2\pi} \right)^2 + \sum_{j=1}^L \left( \left(p_j^{(1)}\right)^2 +  \left( - p_j^{(2)} + 2 p_j^{(1)}\right)^2 \right)  \\ &+   \sum_{j=1}^L \frac{1}{4} \left( -\tilde{p}_{j,j+1}^{(1)} - \frac{\phi_{j+1}^{(1)} - \phi_j^{(1)}}{2\pi} - 2 \tilde{p}_{j,j+1}^{(2)} -2 \frac{\phi_{j+1}^{(2)} - \phi_j^{(2)}}{2\pi} -4 p_{j}^{(2)} + 8 p_{j+1}^{(1)} \right)^2  \\
	&+   \sum_{j=1}^L \frac{1}{4} \left(- \tilde{p}_{j,j+1}^{(2)} - \frac{\phi_{j+1}^{(2)} - \phi_j^{(2)}}{2\pi} + 4 p_{j+1}^{(1)} \right)^2 \,,
\fe
while the constraints, after simplification, become
\ie
	\exp{ \left( 2\pi i q_j^{(1)}  \right)} &= \exp{\left( i  a_j \right)} \,, & \quad \exp{ \left( 2\pi i \tilde{p}_{j-1,j}^{(1)} \right) } &= 1 \,, \\
	\exp{ \left( 2\pi i q_j^{(2)}  \right)} &= 1 \,, & \quad \exp{ \left( 2\pi i \tilde{p}_{j-1,j}^{(2)} \right) } &= 1 \,, \\
	\exp{\left( 2\pi i E_j \right)} &= 1 \,,&
	\tilde{p}^{(1)}_{j,j+1} + E_{j+1} - E_j  &= 0  \,.
\fe

Following the $N=1$ case, we apply the similarity transformation $\exp{ \left( 2\pi i \sum_j E_j q^{(1)}_j \right) }$ to decouple the U(1) gauge fields and get a simpler description of the model. The Hamiltonian in this presentation simplifies to:
\ie
	H_{8,4,-1,2} =&\; \frac{e^2}{8\pi^2} \sum_{j=1}^L  \left(  \phi_j^{(1)} \right)^2 + \sum_{j=1}^L \left( \left(p_j^{(1)}\right)^2 + \left( - p_j^{(2)} + 2 p_j^{(1)}\right)^2 \right)  \\ &+   \sum_{j=1}^L \frac{1}{4} \left(  - 2 \tilde{p}_{j,j+1}^{(2)} -2 \frac{\phi_{j+1}^{(2)} - \phi_j^{(2)}}{2\pi} -4 p_{j}^{(2)} + 8 p_{j+1}^{(1) } - \frac{\phi_{j+1}^{(1)} - \phi_j^{(1)}}{2\pi} \right)^2  \\
	&+   \sum_{j=1}^L \frac{1}{4} \left(- \tilde{p}_{j,j+1}^{(2)} - \frac{\phi_{j+1}^{(2)} - \phi_j^{(2)}}{2\pi} + 4 p_{j+1}^{(1)} \right)^2 \,,
\fe
with constraints
\ie
	\exp{ \left( 2\pi i q_j^{(2)}  \right)} = 1 \,,  \qquad \exp{ \left( 2\pi i \tilde{p}_{j-1,j}^{(2)} \right) } = 1 \,, \label{84-1.gauss}
\fe
and
\ie
	\exp{\left( i  a_j \right)} = 1 \,,  \qquad \exp{\left( 2\pi i E_j \right)} = 1 \,, \qquad \tilde{p}^{(1)}_{j,j+1} = 0  \,.
\fe
Thus the gauge fields $a_j, E_j$ and $\tilde{\phi}^{(1)}_{j,j+1}, \tilde{p}^{(1)}_{j,j+1}$ are completely trivialized and decoupled. Note that the boson $\phi_j^{(1)},p_j^{(1)}$ is noncompact and lives on a tensor product Hilbert space, whereas $\phi_j^{(2)},\tilde{\phi}^{(2)}_{j,j+1}$ is a compact boson whose Hilbert space is the same as the one-flavor modified Villain model.

The lattice chiral gauge theory described above is quadratic and therefore exactly solvable. Specifically, by applying the change of variables introduced in Section \ref{sec:mv.sol}, we can eliminate the Gauss constraints and obtain a genuinely quadratic Hamiltonian. The fields $\phi^{(1)}_j,p^{(1)}_j$ describe a massive boson, akin to the Schwinger boson, of mass of the order $e$, which is coupled to the massless compact boson degrees of freedom ${\phi}_j^{(2)}$ and $\tilde{\phi}_{j,j+1}^{(2)}$. We leave the detailed analysis of this model and its generalizations for future work \cite{WIP}.

\subsection{The fermionic theory}

Using the fermionization procedure, we now construct the 34-50 lattice chiral gauge theory. As explained above, we should fermionically gauge the $\bZ_2^{\ww_2}$ symmetry, which is generated by
\ie
	\prod_{j=1}^L \exp{ \left(  i \pi \, \tilde{p}_{j,j+1}^{(2)} \right) } \,.
\fe

Note that inserting a defect for the $\bZ_2^{\ww_2}$ symmetry amounts to modifying the first Gauss law constraint in \eqref{84-1.gauss} to $e^{2\pi i q_j^{(2)}}=-1$. Following the fermionization procedure reviewed in Appendix \ref{app:fermionization}, we add two Majorana fermions $\psi_j, \tilde{\psi}_j$ at each site and couple them to the $(8,4,-1,2)$ model by replacing the Gauss law constraints in \eqref{84-1.gauss} with
\ie
	\exp{ \left( 2\pi i q_j^{(2)}  \right)} = i \tilde{\psi}_j \psi_j \,, \qquad 	i \psi_j \exp{ \left(  i \pi \, \tilde{p}_{j,j+1}^{(2)} \right) } \tilde{\psi}_{j+1} = 1 \,.
\fe
The second equation is the Gauss law constraint for gauging the $\bZ_2^{\ww_2}$ symmetry.

\section*{Acknowledgments}

I am grateful to Nathan Seiberg for his involvement during the initial stage of this work and for numerous insightful conversations. I also thank Aleksey Cherman, Yichul Choi, Ross Dempsey, Zohar Komargodski, Max Metlitski, Shu-Heng Shao, Stephen Shenker, Nikita Sopenko, Edward Witten, and Wucheng Zhang for valuable discussions, and Shu-Heng Shao for comments on the manuscript. This work was supported in part by the Simons Collaboration on Ultra-Quantum Matter, which is a grant from the Simons Foundation (651444, NS). I also gratefully acknowledge support from the Ambrose Monell Foundation at the Institute for Advanced Study.

\appendix

\section{Exact T-duality on the lattice \label{app:T-duality}}

A nice property of the modified Villain model is its exact T-duality, which we will explore in this section. We first start with the $c=1$ theory.

\subsection{$N=1$}

The $c=1$ compact boson CFT exhibits T-duality, which identifies the theory at radius $R$ with that at radius $1/R$.  As with any duality, T-duality is not unique and is well-defined up to global symmetries of the model.

A choice of T-duality transformation, which we denote by $\ot$, is given by the unitary transformation\footnote{We write $U : \mathcal{O} \mapsto \mathcal{O}'$ as shorthand for  $U \mathcal{O} U^{-1} = \mathcal{O}'$.}
\ie
\ot: \quad \begin{split} \phi_j \mapsto \tilde{\phi}_{j,j+1} \,,& \qquad p_j \mapsto \tilde{p}_{j,j+1} + \frac{\phi_{j+1}-\phi_{j}}{2\pi} \,,  \\
\tilde{\phi}_{j,j+1} \mapsto \phi_{j+1} \,,& \qquad \tilde{p}_{j,j+1} \mapsto p_{j+1} -\frac{\tilde{\phi}_{j+1,j+2}-\tilde{\phi}_{j,j+1}}{2\pi} \,. \end{split} \label{t.duality}
\fe
Crucially, $\ot$ is gauge invariant and preserves Gauss's law constraints. Indeed, it exchanges the Gauss law constraints on sites with those on the links, since
\ie
	\ot : \qquad q_j \mapsto q_{j,j+1} \,, \qquad q_{j,j+1} \mapsto q_{j+1} \,.
\fe
As a result, $\ot$ exchanges the winding and momentum charges, as expected. We can explicitly verify that \eqref{t.duality} implements T-duality since it maps the Hamiltonian at radius $R$ to the dual radius $1/R$:
\begin{equation}
\ot \,H_\mathrm{mV} \, \ot^{-1} = \sum_{j=1}^L \left( \frac{R^2}{2} p_j^2 + \frac{1}{2R^2}  \left( \tilde{p}_{j,j+1} +\frac{{\phi}_{j+1}-\phi_{j}}{2\pi} \right)^2 \right) \,.
\end{equation}

To see that $\ot$ is a well-defined unitary transformation, one may verify that \eqref{t.duality} preserves the operator algebra and, in particular, the canonical commutation relations. A more direct approach is to give an explicit unitary representation of $\ot$. Specifically, we define
\ie
	\ot := \mathbf{T}_{\frac{1}{2}} \mathbf{C} = \mathbf{C}^{-1} \mathbf{T}_{\frac{1}{2}}\,,
\fe
where 
\ie
\mathbf{C} = \exp{\left( i \sum_j \tilde{\phi}_{j,j+1} \frac{\phi_{j+1} - \phi_j}{2\pi} \right)}\,.
\fe
Here $\mathbf{T}_{\frac{1}{2}}$ denotes a half-translation operator, acting as $(\phi_j, p_j) \mapsto (\tilde{\phi}_{j,j+1}, \tilde{p}_{j,j+1})$ and $(\tilde{\phi}_{j,j+1},\tilde{p}_{j,j+1}) \mapsto (\phi_{j+1},p_{j+1})$. 

\subsection{$O(N,N;\mathbb{Z})$ T-duality group}

For the ($N$-flavor) $c=N$ compact boson CFT, the group of T-duality transformations is $O(N,N;\mathbb{Z})$.\footnote{Since the center of $O(N,N;\mathbb{Z})$ acts as the charge conjugation symmetry, the T-duality group that acts faithfully on the moduli space of $c=N$ compact boson CFTs is $PO(N,N;\mathbb{Z})$ rather than $O(N,N;\mathbb{Z})$.} For a review, see, for instance, \cite{Giveon:1994fu,Polchinski:1998rq}. There are three sets of generators for the T-duality group $O(N,N;\mathbb{Z})$. The first one corresponds to doing the T-duality transformation on individual flavors of compact bosons, which we have already discussed above. Another is the duality group $GL(N,\mathbb{Z})$ of large diffeomorphisms of $T^N$. The final set corresponds to the integer shift of the antisymmetric $B$-field.

We identify $O(N,N;\mathbb{Z})$ as the group of $2N \times 2N$ integer matrices $\mathcal{O}$ satisfying
\ie
	\mathcal{O}^T \begin{pmatrix}
0 & \mathbbm{1}_{N} \\ 
\mathbbm{1}_{N} & 0
\end{pmatrix} \mathcal{O} = \begin{pmatrix}
0 & \mathbbm{1}_{N} \\ 
\mathbbm{1}_{N} & 0
\end{pmatrix} \,,
\fe
which acts on the momentum and winding charge vectors $(\vec{n}_\mm, \vec{n}_\ww) \in \mathbb{Z}^N \oplus \mathbb{Z}^N$. In terms of such matrices, the three sets of generators mentioned above correspond to the following:
\begin{enumerate}
\item T-duality on the $I$-th compact boson component is
\ie
	\mathcal{O}_I =\begin{pmatrix}
\mathbbm{1}_N - P_I & P_I \\ 
P_I & \mathbbm{1}_N - P_I
\end{pmatrix} \,, \qquad \left(P_I\right)_{JK} = \delta_{J,I} \delta_{K,I} ~,
\fe
with $P_I$ the $N\times N$ rank-one projector onto the $I$-th basis vector.
\item Large diffeomorphisms of $T^N$
\ie
	\mathcal{O}_S =\begin{pmatrix}
S & 0 \\ 
0& (S^{-1})^{T} 
\end{pmatrix} \,, \qquad S \in GL(N, \mathbb{Z})
\fe 
\item Integer shift of the antisymmetric $B$-field is implemented by
\ie
	\mathcal{O}_\Theta =\begin{pmatrix}
\mathbbm{1}_N & \Theta \\ 
0 & \mathbbm{1}_N 
\end{pmatrix} \,,
\fe
for $\Theta$ an $N\times N $ antisymmetric integer matrix.
\end{enumerate}

Given a matrix $\mathcal{O} \in O(N,N; \mathbb{Z})$, the corresponding T-duality transformation $\mathcal{T}_\mathcal{O}$ acts on the momentum and winding charges of the $N$-flavor modified Villain theory as
\ie
	\mathcal{T}_\mathcal{O} \left( \vec{n} \cdot \vec{Q} \right) {\mathcal{T}_\mathcal{O}}^{-1} = \vec{\mathcal{O} n} \cdot \vec{Q}\,,
\fe
for any integer vector $\vec{n} = (\vec{n}_\mm , \vec{n}_\ww) \in \bZ^N \oplus \bZ^N$. In other words, we have
\ie
	\mathcal{T}_\mathcal{O} \begin{pmatrix}
\vec{Q}_\mm \\ 
\vec{Q}_\ww
\end{pmatrix}  {\mathcal{T}_\mathcal{O}}^{-1} = \mathcal{O}^{T} \cdot \begin{pmatrix}
\vec{Q}_\mm \\ 
\vec{Q}_\ww
\end{pmatrix}  \qquad \text{and} \qquad \mathcal{T}_{\mathcal{O} \mathcal{O}'} = \mathcal{T}_{\mathcal{O}}\mathcal{T}_{\mathcal{O}'} \,.
\fe

\subsection{$N=2$}

Here, we find these generators explicitly for the case of $c=N=2$ modified Villain theory.

\paragraph{T-duality on individual bosons:} The first set of generators corresponds to T-duality transformations $\ot_1$ and $\ot_2$ on individual flavors of compact boson. As we discussed above, their action on local operators is
\ie
\ot_I: \quad \begin{split} \phi_j^{(I)} \mapsto \tilde{\phi}_{j,j+1}^{(I)} \,,& \qquad p_j^{(I)} \mapsto \tilde{p}_{j,j+1}^{(I)} + \frac{\phi_{j+1}^{(I)}-\phi_{j}^{(I)}}{2\pi} \,,  \\
\tilde{\phi}_{j,j+1}^{(I)} \mapsto \phi_{j+1}^{(I)} \,,& \qquad \tilde{p}_{j,j+1}^{(I)} \mapsto p_{j+1}^{(I)} -\frac{\tilde{\phi}_{j+1,j+2}^{(I)}-\tilde{\phi}_{j,j+1}^{(I)}}{2\pi} \,. \end{split}
\fe

\paragraph{Large diffeomorphisms of $T^2$:} The second type corresponds to $GL(2,\mathbb{Z})$ transformations acting on the two flavors. The straightforward transformations correspond to $GL(2,\mathbb{Z})$ matrices
\ie
	\begin{pmatrix}
\pm 1 & 0 \\ 
0 & \pm 1
\end{pmatrix} \qquad \text{and} \qquad  \begin{pmatrix}
0 & 1 \\ 
1 & 0
\end{pmatrix} \,,
\fe
which, respectively, act as charge conjugation on individual bosons and by permuting the two flavors. The only non-trivial generator corresponds to 
\ie
	\mathcal{O}_\mathsf{T} =\begin{pmatrix}
\mathsf{T} & 0 \\ 
0& (\mathsf{T}^{-1})^{T} 
\end{pmatrix} \,, \qquad \mathsf{T} = \begin{pmatrix}
 1 & 1 \\ 
0 &  1
\end{pmatrix} \,,
\fe
which is implemented by the unitary transformation
\ie
	\ot_\mathsf{T} = \exp{\left[ i \sum_{j=1}^L \left( -\phi_j^{(2)} p_j^{(1)} +\tilde{\phi}^{(1)}_{j,j+1} \tilde{p}^{(2)}_{j,j+1}  \right) \right]} \,.
\fe
Its action on local operators is
\ie
\ot_\mathsf{T}: \quad \begin{split} \phi_j^{(1)} \mapsto \phi_j^{(1)} - \phi_j^{(2)} \,,& \qquad p_j^{(2)} \mapsto p_j^{(2)} + p_j^{(1)} \,,  \\
\tilde{\phi}_{j,j+1}^{(2)} \mapsto \tilde{\phi}_{j,j+1}^{(2)} + \tilde{\phi}_{j,j+1}^{(1)} \,,& \qquad \tilde{p}_{j,j+1}^{(1)} \mapsto \tilde{p}_{j,j+1}^{(1)} - \tilde{p}_{j,j+1}^{(2)} \,, \end{split}
\fe
which indeed satisfy
\ie
	\mathcal{T}_\mathsf{T} \begin{pmatrix} 
\vec{Q}_\mm \\[10pt]
\vec{Q}_\ww
\end{pmatrix}  {\mathcal{T}_\mathsf{T}}^{-1} = (\mathcal{O}_\mathsf{T})^{T} \cdot \begin{pmatrix}
\vec{Q}_\mm \\[10pt]
\vec{Q}_\ww
\end{pmatrix} = \begin{pmatrix}
Q^{(1)}_\mm  \\ 
Q^{(2)}_\mm + Q^{(1)}_\mm\\
Q^{(1)}_\ww - Q^{(2)}_\ww \\
Q^{(2)}_\ww
\end{pmatrix} \,.
\fe

\paragraph{$B$-field integer shift:} Finally, the third type of T-duality transformations corresponds to
\ie
	\mathcal{O}_\varepsilon =\begin{pmatrix}
\mathbbm{1}_2 & \varepsilon \\ 
0& \mathbbm{1}_2 
\end{pmatrix} \,, \qquad \varepsilon = \begin{pmatrix}
 0 & 1 \\ 
-1 &  0
\end{pmatrix} \,,
\fe
which is implemented by
\ie
	\ot_\varepsilon &= \exp{\left[ i \sum_{j=1}^L \left(  \tilde{\phi}^{(1)}_{j,j+1} \bigg( p^{(2)}_{j+1} - \frac{ \tilde{\phi}^{(2)}_{j+1,j+2} - \tilde{\phi}^{(2)}_{j,j+1} }{2\pi} \bigg)   - p_j^{(1)} \tilde{\phi}_{j,j+1}^{(2)} \right)  \right]} \\
	&= \exp{\left[ i \sum_{j=1}^L \left( - \tilde{\phi}^{(2)}_{j,j+1} \bigg( p^{(1)}_{j} - \frac{ \tilde{\phi}^{(1)}_{j,j+1} - \tilde{\phi}^{(1)}_{j-1,j} }{2\pi} \bigg) +  p_{j+1}^{(2)} \tilde{\phi}_{j,j+1}^{(1)}   \right) \right]} \,.
\fe
Its action on local operators is
\ie
\ot_\varepsilon: \quad \begin{split} {\phi}_{j}^{(1)} \mapsto {\phi}_{j}^{(1)} - \tilde{\phi}_{j,j+1}^{(2)} \,,& \qquad \tilde{p}_{j,j+1}^{(1)} \mapsto \tilde{p}_{j,j+1}^{(1)} - \left( p^{(2)}_{j+1} - \frac{ \tilde{\phi}^{(2)}_{j+1,j+2} - \tilde{\phi}^{(2)}_{j,j+1} }{2\pi} \right) \,,  \\
{\phi}_{j}^{(2)} \mapsto {\phi}_{j}^{(2)} + \tilde{\phi}_{j-1,j}^{(1)} \,,& \qquad \tilde{p}_{j,j+1}^{(2)} \mapsto \tilde{p}_{j,j+1}^{(2)} + \left( {p}^{(1)}_{j} - \frac{ \tilde{\phi}^{(1)}_{j,j+1} - \tilde{\phi}^{(1)}_{j-1,j} }{2\pi} \right) \,, \end{split} 
\fe
while it leaves $p_j^{(I)}$ and $\tilde{\phi}_{j,j+1}^{(I)}$ invariant. One can verify that it is gauge-invariant since it leaves $q_{j}^{(I)}$ invariant and
\ie
	\ot_\varepsilon: \quad q_{j,j+1}^{(1)} \mapsto q_{j,j+1}^{(1)} - q_{j+1}^{(2)} \,, \qquad q_{j,j+1}^{(2)} \mapsto q_{j,j+1}^{(2)} + q_{j}^{(1)} \,.
\fe
As a result, it satisfies the expected T-duality transformation
\ie
	\mathcal{T}_\varepsilon \begin{pmatrix} 
\vec{Q}_\mm \\[10pt]
\vec{Q}_\ww
\end{pmatrix}  {\mathcal{T}_\varepsilon}^{-1} = (\mathcal{O}_\varepsilon)^T  \cdot \begin{pmatrix}
\vec{Q}_\mm \\[10pt]
\vec{Q}_\ww
\end{pmatrix} = \begin{pmatrix}
Q^{(1)}_\mm  \\ 
Q^{(2)}_\mm \\
Q^{(1)}_\ww - Q^{(2)}_\mm\\
Q^{(2)}_\ww + Q^{(1)}_\mm
\end{pmatrix} \,.
\fe

The T-duality associated with the transpose matrix $(\mathcal{O}_\varepsilon)^T$ is implemented by
\ie
	\ot^T_\varepsilon &= \exp{\left[ i \sum_{j=1}^L \left( -\phi_j^{(1)} \tilde{p}_{j,j+1}^{(2)} + \bigg( \tilde{p}^{(1)}_{j,j+1} + \frac{ {\phi}^{(1)}_{j+1} - {\phi}^{(1)}_{j}}{2\pi} \bigg) \phi^{(2)}_{j+1}  \right) \right]} \\
	&= \exp{\left[ i \sum_{j=1}^L \left( + \tilde{p}_{j,j+1}^{(1)} \phi_{j+1}^{(2)} - \phi^{(1)}_{j} \bigg( \tilde{p}^{(2)}_{j,j+1} + \frac{ {\phi}^{(2)}_{j+1} - {\phi}^{(2)}_{j}}{2\pi} \bigg)   \right) \right]} \,.
\fe
Its action on local operators is
\ie
\ot_\varepsilon^T: \quad \begin{split} \tilde{\phi}_{j,j+1}^{(1)} \mapsto \tilde{\phi}_{j,j+1}^{(1)} + {\phi}_{j+1}^{(2)} \,,& \qquad p_j^{(1)} \mapsto p_j^{(1)} + \left( \tilde{p}^{(2)}_{j,j+1} + \frac{ {\phi}^{(2)}_{j+1} - {\phi}^{(2)}_{j}}{2\pi} \right) \,,  \\
\tilde{\phi}_{j,j+1}^{(2)} \mapsto \tilde{\phi}_{j,j+1}^{(2)} - {\phi}_{j}^{(1)} \,,& \qquad p_j^{(2)} \mapsto p_j^{(2)} - \left( \tilde{p}^{(1)}_{j-1,j} + \frac{ {\phi}^{(1)}_{j} - {\phi}^{(1)}_{j-1}}{2\pi} \right) \,, \end{split} 
\fe
while it leaves $\phi_j^{(I)}$ and $\tilde{p}_{j,j+1}^{(I)}$ invariant. One can verify that it is gauge-invariant since it leaves $q_{j,j+1}^{(I)}$ invariant and
\ie
	\ot^T_\varepsilon: \quad q_j^{(1)} \mapsto q_j^{(1)} +  q_{j,j+1}^{(2)} \,, \qquad q_j^{(2)} \mapsto q_j^{(2)} -  q_{j-1,j}^{(1)} \,.
\fe
As a result, it satisfies the expected T-duality transformation
\ie
	\mathcal{T}^T_\varepsilon \begin{pmatrix} 
\vec{Q}_\mm \\[10pt]
\vec{Q}_\ww
\end{pmatrix}  (\mathcal{T}_\varepsilon^T)^{-1} = \mathcal{O}_\varepsilon \cdot \begin{pmatrix}
\vec{Q}_\mm \\[10pt]
\vec{Q}_\ww
\end{pmatrix} = \begin{pmatrix}
Q^{(1)}_\mm + Q^{(2)}_\ww \\ 
Q^{(2)}_\mm - Q^{(1)}_\ww\\
Q^{(1)}_\ww \\
Q^{(2)}_\ww
\end{pmatrix} \,.
\fe

\section{Fermionization on the lattice \label{app:fermionization}}

Here, we review the fermionization on the lattice following \cite{Seifnashri:2023dpa, Pace:2024oys}.\footnote{For recent reviews of bosonization in the continuum, see \cite{Gaiotto:2015zta, Tachikawa2018Topological, Karch:2019lnn} and the references therein.} Namely, we discuss how to do fermionic gauging of an anomaly-free $\bZ_2$ symmetry in a bosonic 1+1d lattice model and obtain a fermionic theory. We employ a microscopic description of fermionization, where gauging is achieved by adding fermionic gauge fields and imposing Gauss's law constraints. For a global perspective on bosonization/fermionization on the lattice, see \cite{Seiberg:2023cdc}.

For simplicity, we focus on gauging a $\bZ_2$ symmetry of the form
\ie
	U = \prod_{j=1}^L U_j \,,
\fe
where the local factors $U_j$ mutually commute with each other and satisfy $U_j^2=1$. All $\bZ_2$ symmetries of this form are anomaly-free. Given a $\bZ_2$-symmetric Hamiltonian $H$, to do the fermionic gauging, we first tensor it with an \emph{invertible} fermionic theory and gauge the diagonal $\bZ_2$ symmetry $\mathrm{diag}(\bZ_2 \times \bZ_2^{\psi})$, where $\bZ_2^{\psi}$ is the symmetry whose defects corresponds to the transparent fermion line (worldline of a massive fermion) in the trivial fermionic theory.

Specifically, we start with stacked theory
\ie
	H_\lambda = H - i\lambda \sum_{j=1}^L  \tilde{\psi}_{j,j+1} \psi_{j,j+1} \,.
\fe
Here, $\psi_{j,j+1}$ and $\tilde{\psi}_{j,j+1}$ are Majorana fermions satisfying the anticommutation relations $\{ \psi_{j,j+1} , \psi_{j',j'+1} \} = 2 \delta_{j,j'}$ and $\{ \tilde{\psi}_{j,j+1} , \tilde{\psi}_{j',j'+1} \} = 2 \delta_{j,j'}$. In the limit $\lambda \to \infty$, the fermionic theory becomes completely trivial, and in particular invertible. Now, we gauge the diagonal $\bZ_2$ symmetry generated by
\ie
	\eta = \prod_{j=1}^L \left( i \psi_{j-1,j} U_j \tilde{\psi}_{j,j+1} \right)\,.
\fe
The Hamiltonian of the gauged theory is
\ie
	\tilde{H} = H_{Z_{1,2} ; Z_{2,3} ; \cdots ; Z_{L,1}} - i\lambda \sum_{j=1}^L \tilde{\psi}_{j,j+1} Z
	_{j,j+1} \psi_{j,j+1}\,, \label{app.g.hamiltonian}
\fe
where the Pauli operator $Z_{j,j+1}$ denotes a $\bZ_2$ gauge field on link $(j,j+1)$. The defect Hamiltonian $H_{z_{1,2} ; z_{2,3} ; \cdots ; z_{L,1}}$ is the Hamiltonian of the original theory coupled to the gauge field configuration $z_{j,j+1}$ \cite{Seifnashri:2023dpa}. Gauss's law on site $j$ is
\ie
	X_{j-1,j} \left( i \psi_{j-1,j} U_j \tilde{\psi}_{j,j+1} \right) X_{j,j+1} = 1 \,. \label{app.equation.gauss}
\fe
The final answer is the $\lambda \to +\infty$ limit of this theory. In this limit, we drop the second term in \eqref{app.g.hamiltonian} and instead impose the constraints
\ie
	Z_{j,j+1} = i  \tilde{\psi}_{j,j+1} \psi_{j,j+1}\,.
\fe
By doing a change of variables, we can completely get rid of the $X_{j,j+1}, Z_{j,j+1}$ degrees of freedom, which effectively amounts to setting $X_{j,j+1} = 1$ in \eqref{app.equation.gauss}. More precisely, this corresponds to substitute $X,Z$ and $\psi, \tilde{\psi}$ by the gauge invariant Majorana fermions
\ie
	\psi'_{j,j+1} = X_{j,j+1} \psi_{j,j+1} \qquad \text{and} \qquad \tilde{\psi}'_{j,j+1} = X_{j,j+1} \tilde{\psi}_{j,j+1}
\fe

In summary, fermionic gauging corresponds to adding two Majorana fermions per link, where the $\bZ_2$ gauge field at link $(j,j+1)$ is given by $i \psi_{j,j+1} \tilde{\psi}_{j,j+1}$ and Gauss's law constraints are given by
\ie
	i \psi_{j-1,j} U_j \tilde{\psi}_{j,j+1} = 1\,.
\fe

\subsection{The Ising model}

We now apply the fermionization procedure described above to the transverse-field Ising model, whose Hamiltonian is
\ie
	H_\mathrm{Ising} = - \sum_{j=1}^L \left( Z_j Z_{j+1} + h X_j \right)\,.
\fe
Fermionically gauging the $\bZ_2$ symmetry generated by $\prod_j X_j$ leads to the Hamiltonian
\ie
	H_\mathrm{fermionic} = - \sum_{j=1}^L \left( Z_j ( i \tilde{\psi}_{j,j+1} \psi_{j,j+1}  ) Z_{j+1} + h X_j \right)\,,
\fe
subject to Gauss's law constraints 
\ie
	i\psi_{j-1,j} X_j \tilde{\psi}_{j,j+1} = 1\,.
\fe

As discussed above, we use Gauss's law constraints and perform a change of variables to eliminate the spin degrees of freedom $X_j$ and $Z_j$.  As expected, this yields the following free fermion Hamiltonian
\ie
	H_\mathrm{fermionic} = -i \sum_{j=1}^L \left( \tilde{\psi}_{j-1,j} \psi_{j-1,j}  + h \psi_{j-1,j} \tilde{\psi}_{j,j+1} \right)\,.
\fe

\bibliographystyle{JHEP}
\bibliography{refs}
\end{document}